\documentclass[pre,reprint,amsmath]{revtex4-1}
\usepackage[utf8]{inputenc}
\usepackage{graphicx}
\usepackage[colorlinks=true,linkcolor=blue,citecolor=blue]{hyperref}

\begin{document}
\title{Swift-Hohenberg equation with third order dispersion for optical fiber resonators}
\author{A. Hariz}
\author{L. Bahloul}
\author{L. Cherbi}

\affiliation{Laboratoire d'instrumentation, Université des Sciences et de la Technologie Houari Boumediene (USTHB), Bab Ezzouar, Algeria}

\author{K. Panajotov}

\affiliation{Vrije Universiteit Brussel, Department of Applied Physics \& Photonics,
Pleinlaan 2, B-1050 Brussels, Belgium}

\affiliation{Institute of Solid State Physics,  72 Tzarigradsko Chaussee Blvd.,  1784 Sofia,  Bulgaria}
\author{M. Clerc}
\author{M.A. Ferr\'e}
\affiliation{Departamento de Fisica {and Millennium Institute for Research in Optics}, FCFM, Universidad de Chile, 
Casilla 487-3, Santiago, Chile}
\author{B. Kostet}
\author{E. Averlant}
\author{M. Tlidi}
\affiliation{Faculté des Sciences, Université Libre de Bruxelles (ULB), P.231, Campus Plaine, B-1050 Brussels, Belgium}
\begin{abstract}
We investigate the dynamics of a ring cavity made of photonic crystal fiber and driven by a coherent beam
 working near to the resonant frequency of the cavity.  
By  means of a multiple-scale reduction of the Lugiato-Lefever equation with high order dispersion, we show that the dynamics of this 
optical device, when operating close to the critical point associated with bistability, is captured by a  
real order parameter equation in the form of a generalized  Swift-Hohenberg equation. 
A Swift-Hohenberg equation has been derived for several areas of nonlinear science 
such as chemistry, biology, ecology, optics, and laser physics. 
However,  the peculiarity of the obtained generalized Swift-Hohenberg equation for photonic crystal 
fiber resonators is that it possesses a third-order dispersion. Based on a weakly 
nonlinear analysis in the vicinity of the modulational instability threshold, 
we characterize the motion of dissipative structures by estimating their propagation speed.  
Finally, we numerically investigate the formation of moving temporal localized structures often called cavity solitons.
\end{abstract}
\maketitle
\section{Introduction}
The control of linear and nonlinear properties of Photonic Crystal Fibers (PCF) has led to several applications in optoelectronics, sensing, and laser science (see recent overview on this issue~\cite{Markos_17} and references therein). The improvement of the fabrication capabilities of high-quality integrated microstructure
resonators such as ring, micro-ring, micro-disc, or Fabry-Perot cavities are drawing
considerable attention from both fundamental and applied points of view. Among those, an important class is the photonic crystal fiber resonators where high-order
chromatic dispersion  could play an important role in the dynamics  \cite{Cavalcanti,Joly}, particularly
in relation with supercontinuum generation \cite{Yulin,Dudley}. On the other hand, driven optical microcavities are widely used for the generation of optical frequency combs. They can be modeled by the Lugiato-Lefever equation \cite{Lugiato1987} that possesses solutions in the form of localized structures (LSs) \cite{Scroggie,Leo}. Optical
frequency-combs generated in high-Q Kerr resonators \cite{Kippenberg} are in fact the spectral content of the stable temporal pattern occurring in the
cavity. Among the possible dissipative structures, the temporal localized structures often called dissipative solitons appear in the form of a stable single pulse on top of a low background. They have been theoretically predicted in \cite{Scroggie} and experimentally observed in \cite{Leo}. 
The link between temporal localized structures and  Kerr comb (KC) generation in high-Q resonators has motivated further interest in this issue. 

\begin{figure}[b]
\includegraphics[width=.45\textwidth]{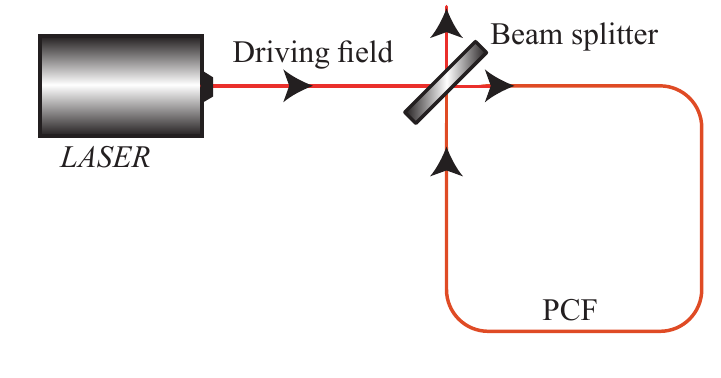}
\caption{\label{fig:LL}A schematic representation of a ring cavity filled with a photonic crystal fiber (PCF) and driven by a coherent beam.}
\end{figure}

In free propagation, the envelope of the light inside a single-mode optical fiber is described by 
the nonlinear Schrödinger equation, in which the propagation constant is expanded 
in a Taylor series up to fourth order in the frequency domain. When inserting a 
photonic crystal fibers in a cavity which is driven by a continuous beam, the light inside the fiber is coherently superimposed with the input beam at each round trip. The nonlinear Schrödinger equation supplemented by the cavity boundary condition leads to a generalized Lugiato-Lefever equation~\cite{Tlidi_07}. The inclusion of the fourth order dispersion allows the modulational instability (MI) to have a finite domain of existence delimited by two pump power values \cite{Tlidi_07}, as well as to stabilize  dark temporal LSs \cite{Tlidi3,ALyes,Bahloul_PTRS}. 
In the absence of fourth order dispersion, with only second or/and third orders of dispersion, fronts interaction leads to the formation of moving LS in a regime far from any modulational instability \cite{Fronts}. An analytical study of the interaction between LSs under the action of Cherenkov radiation or dispersive waves has been conducted in \cite{Andrei_18}. In addition, it has been shown that  the interference between
the dispersive waves emitted by the two interacting LSs produces
an oscillating pattern responsible for the stabilization of the bound states. A derivation of equations governing the time
evolution of the position of two well-separated LSs interacting
weakly via their exponentially decaying tails have been also derived \cite{Andrei_18}. 

In this paper, we derive a generalized Swift-Hohenberg equation with third order dispersion
 describing the evolution of pulses 
propagating in a photonic crystal fiber resonator. This reduction is performed for small-frequency 
modes and close to a second-order critical point marking the onset 
of a hysteresis loop (nascent bistability). 
The dimensionless generalized real Swift-Hohenberg equation (SHE)  reads
\begin{equation}
\frac{\partial \phi}{\partial t} =y+c\phi -\phi^{3}+\beta\frac{\partial ^{2}\phi }{\partial \tau^{2}} +\beta^{\prime}\frac{\partial ^{3}\phi }{\partial \tau ^{3}}-\frac{\partial^{4}\phi }{\partial \tau ^{ 4}}.
\label{Eq-1}
 \end{equation}
Here $\phi(t,\tau)$ is the deviation of the electric field from its value at the onset of bistability. The variable  $\phi$ is a scalar real field.
The time variable $t$ corresponds to the slow evolution of $\phi$ over successive
round-trips. $\tau$ accounts for the fast dynamics that describes
how the electric field envelope changes along the fiber. 
The parameters $y$ and $c$ are deviations of the cooperativity parameter and the amplitude of the injected field from their critical values at the onset of the modulational  instability, respectively. The coefficients $\beta$ and $\beta^{\prime}$ account for second and third order chromatic dispersion, respectively. Without loss of generality, we rescale the fourth order dispersion coefficient  and the cubic coefficient to unity. The third and the fourth orders of dispersion are usually neglected in fiber cavity models \cite{Lugiato1987}. However, the dispersion characteristics of photonic crystal resonators impose to consider high order dispersion \cite{Biancalana_03,Biancalana_04}.

The paper is organized as follows. After an introduction, we present the Lugiato-Lefever equation with high order dispersion and we perform a linear stability analysis of the homogeneous steady states in section II. In section III, we derive a generalized real Swift-Hohenberg equation. A weakly nonlinear analysis of the real Swift-Hohenberg equation is presented in section VI. In this section, we also present a derivation of their linear and nonlinear velocities. Moving temporal localized structures with a single peak or more peaks are demonstrated in section V.  In the limit of small third-order dispersion, an analytical expression for the speed of the localized structure is provided in the section VI . We conclude in section VII.

\section{Lugiato-Lefever equation with higher order dispersions}
We consider an optical cavity with a length $L$ filled with a photonic crystal fiber, and synchronously pumped by a coherent injected beam, as described in Fig.~(\ref{fig:LL}). A continuous wave $E_{i}$ is injected into the cavity by means of a beam splitter. The field $E$ propagates inside the fiber and experiences dispersion, Kerr effect, and dissipation. The linear phase shift accumulated during one  cavity round-trip is denoted by $\Phi_{0}$. The intensity mirror transmissivity (reflectivity) is $T^{2}$ ($R^{2}$). 
The second, third and fourth order dispersion terms are, respectively, $\beta_{2,3,4}$. $\tau$ is the time in the reference frame moving at the group
velocity of the light describing the fast evolution of the field envelope within the cavity; $\tau \equiv \tau \left( T^{2}/L\right) ^{1/2}$. The time $t$, describes the slow evolution of the field envelope between
two consecutive cavity round-trips, and is  scaled such that
the decay rate  is unity, i.e., $t\equiv tT^2/2t_{r}$, where $t_r$ is the round-trip time. The normalized  intracavity field is $E \equiv E\sqrt{
2\gamma L/T^{2}}$, and the injected field is $S=2/T\left( 2\gamma L/T^{2}\right) ^{1/2}E_{i}$, with  $\gamma$ the nonlinear coefficient. The normalized cavity detuning is  $\theta =2\Phi _{0}/T^{2}$. We also replace the coefficients $\beta_{2,3,4}$ by $\beta$ and $\beta^\prime$ through the relations $\partial/\partial \tau \equiv \beta_4^{-1/4} \partial/\partial \tau$, $\beta = \beta_2 / \sqrt{\beta_4}$ and $\beta^\prime = \beta_3 / \beta_4^{3/4}$.

In its dimensionless form, the generalized Lugiato-Lefever equation studied in~\cite{Tlidi_07}, reads
\begin{eqnarray}
\frac{\partial E}{\partial t} =& & S-(1+i\theta )E +i\left|E \right| ^{2}E-i\beta\frac{\partial ^{2}E }{\partial \tau^{2}} \nonumber \\
&+&\beta{'}\frac{\partial ^{3}E }{\partial \tau ^{3}}+i\frac{\partial^{4}E }{\partial \tau ^{ 4}}. 
\label{eq:LL}
\end{eqnarray}

The homogeneous stationary solutions (HSSs), $E_s$, of Eq.~(\ref{eq:LL}) are described by $S^2=I_s[1+(\theta-I_s)^2],$ and $I_s=|E_s|^2$. 
This system will exhibit a bistable behavior for $\theta>\theta_c=\sqrt3$, 
and a monostable behavior for $\theta \leq \theta_c$. 
The linear stability of the homogeneous steady states has been performed  in~\cite{Tlidi_07}. 
The presence of a fourth order dispersion gives rise to (i) a degenerate modulational instability where two separate frequencies simultaneously appear, and (ii) appearance of a second MI that stabilizes the high intensity regime  ~\cite{Tlidi_07}.  The linear stability analysis of the homogeneous solutions with respect to finite frequency perturbation of the form $\exp(i\Omega\tau+\lambda t)$ yields
\begin{equation}
\lambda_{\pm}=-1+i\beta^\prime \Omega^3 \pm\sqrt{I_s^2-(\theta-2I_s-\beta \Omega^2-\Omega^4)^2}.  \label{eigenvalues}
\end{equation}
This dispersion relation through the conditions $\partial \lambda/\partial \Omega=\partial^2  \lambda/\partial \Omega ^2=0$ yields expressions for the critical frequencies at the first MI bifurcation which are degenerate
\begin{equation}
\Omega_{l,u}^{2}=\frac{-\beta\pm\sqrt{\beta^{2}+4(\theta-2I_s\pm \sqrt{I_s^2-1})}}{2}.     
\label{eq:omega_LU}
\end{equation}
These two frequencies ($\Omega_{l}$ and $\Omega_{u}$) are simultaneously and spontaneously generated at the primary threshold $I_s=I_{1m}=1$. When the two critical frequencies $\Omega_{l,u}$ are close to each other, it has been shown that an intrinsic beating frequencies appear $\Omega_{l} \pm\Omega_{u}$ \cite{Kozyreff}. Besides this first degenerate modulational instability, the fourth order dispersion allows for the stabilization of the high intensity regime by creating another MI. The critical value of the frequency at the upper bifurcation point $I_{2m}$ is given by
$\Omega_{c}^{2}=-\beta/4$. It is then possible to restabilize the stationary state by driving the system to the large intensity regime ($I>I_{2m}$).

\section{A generalized Swift-Hohenberg equation for photonic crystal fiber resonators}
\label{sec:linana}
\subsection{Derivation of a generilized Swift-Hohenberg equation}
The purpose of this section is to present the derivation of the  generalized 
Swift-Hohenberg for a photonic crystal fiber resonator. For this purpose, we use multiple scale method. 
To this aim, we
explore the space time dynamics in the vicinity of the critical point associated with the  nascent bistability. More precisely, for $\theta=\theta_c$, there exists a second order critical point marking the onset of an hysteresis loop. This transition point is defined by
$\partial S/\partial I_s=\partial^2S/\partial I_s^2=0$. The coordinate of this critical point are:  
\begin{equation}
E_c=\left(\frac{3}{4}-i\frac{\sqrt{3}}{4}\right)S_{c}, { \mbox{ and  }} S_{c}^{2}=\frac{8\sqrt{3}}{9}.
\label{CritiINT}
\end{equation}
We decompose the electrical field in real and imaginary parts as $E=X_1+iX_2$, replace it in Eq.~(\ref{eq:LL}) and get
\begin{eqnarray}
\frac{\partial X_1}{\partial t}&=&S-X_1 + \theta X_2-X_2X_1^2-X_2^3  \\ \nonumber 
&+&\beta\frac{\partial ^{2}X_2 }{\partial \tau^{2}}+\beta{'}\frac{\partial ^{3}X_1 }{\partial \tau ^{3}}-\frac{\partial^{4}X_2 }{\partial \tau ^{ 4}},
\label{eq:LLX1}\\
\frac{\partial X_2}{\partial t}&=&-X_2 - \theta X_1+X_1X_2^2+X_1^3\\ \nonumber&+&\beta\frac{\partial ^{2}X_2 }{\partial \tau^{2}} +\beta{'}\frac{\partial ^{3}X_1 }{\partial \tau ^{3}}-\frac{\partial^{4}X_2 }{\partial \tau ^{ 4}}.\label{eq:LLX2}
\end{eqnarray}
Introducing the excess variables $U$ and $V$: $X_1(\tau,t)=x_{1s}+U(\tau,t)$ and  $X_2(\tau,t)=x_{2s}+V(\tau,t)$ with $x_{1s}$ and  $x_{2s}$, respectively, the real and imaginary parts of the homogeneous stationary solution given by
\begin{eqnarray}
S-x_{1s} + \theta x_{2s}-x_{2s}x_{1s}^2-x_{2s}^3 &=& 0\\
-x_{2s} - \theta x_{1s}+x_{1s}x_{2s}^2+x_{1s}^3 &=& 0.
\end{eqnarray}
By solving these algebraic equations and using Eq.~(\ref{CritiINT}), the critical values of $x_{1s}$ and $ x_{2s}$ are found to be $x_{1c}=3S_{c}/4$ and  $x_{2c}=-\sqrt{3}S_{c}/4$. In order to explore the vicinity of the nascent hysteresis, we introduce a small parameter  $\epsilon$ which measures the distance from the critical point associated with bistability 
\begin{equation}
\theta=\sqrt{3}+\epsilon^{2}\delta.
\end{equation}
Indeed, $\delta$ accounts for the separation of detuning with respect to critical one.
We then expand in powers of $\epsilon$  the excess variables and the injected field as
\begin{eqnarray}
U&=&\epsilon u_{0}+\epsilon^{2} u_{1}+\epsilon^{3}  u_{2}+...\\
V&=&\epsilon v_{0}+\epsilon^{2}  v_{1}+\epsilon^{3}  v_{2}+...\\
S &=&S_{c}+\varepsilon S_{1}+\varepsilon^2 S_{2}+\varepsilon^3 S_{3}+...,
\end{eqnarray}
and also introduce the slow and the fast time scales, $t  \equiv t / \varepsilon^2 $ and  $\tau  \equiv \tau/ \sqrt{\varepsilon}$.
A preliminary analysis indicates that we
need to consider that $\beta$ is small $\beta \equiv \varepsilon \beta$. 
This is because we consider a low frequency (or large period) regime. In this way, 
the MI instability threshold is close to the critical point. We now replace 
all the above scalings and expansions in Eq.~(\ref{eq:LLX1}) and Eq.~(\ref{eq:LLX2}), and make an
expansion in series of $\varepsilon $ up to the order $\varepsilon ^{3}.$
To $O\left( \varepsilon \right)$, one gets
\begin{equation}
\left( \begin{array}{c}S_{1}\\
0
\end{array}\right)= \begin{bmatrix}
    1+2x_{1c}x_{2c} & x_{1c}^2+3 x_{2c}^2-\theta_c\\
    -x_{2c}^2-3 x_{1c}^2+\theta_c & 1-2x_{1c}x_{2c}
  \end{bmatrix}\left( \begin{array}{c}u_0\\
v_0
\end{array}\right)\nonumber \\  
\label{eq:fluc}
\end{equation}
By replacing the values of $\theta_c$ and $x_{1c,2c}$ in these equations, the solvability 
condition yields: $S_1=0$ and $u_0(\tau,t)=\sqrt{3}v_0(\tau,t)$.
To the next order $O\left( \varepsilon^2 \right) $, we obtain
\begin{eqnarray}
\begin{bmatrix}
    1+2x_{1c}x_{2c} & x_{1c}^2+3 x_{2c}^2-\theta_c\\
    -x_{2c}^2-3 x_{1c}^2+\theta_c & 1-2x_{1c}x_{2c}
  \end{bmatrix}\left( \begin{array}{c}u_1\\
v_1
\end{array}\right) \nonumber \\  
=
\left( \begin{array}{c}s_{2}+x_{2c}\delta\\
-x_{1c}\delta
\end{array}\right)
+
\left(\begin{array}{c} -x_{2c}u_{0}^2-3x_{2c}v_0^2-2x_{1c}u_0v_0\nonumber \\
 x_{1c}v_{0}^2+3x_{1c}u_0^2+2x_{2c}u_0v_0
\end{array}\right)\nonumber 
\label{eq:fluc}
\end{eqnarray}
In these equations, we replace the values of $\theta_c$, $x_{1c,2c}$,  $s_1=0$, $u_0(\tau,t)=\sqrt{3}v_0(\tau,t)$, and the solvability condition leading to
\begin{eqnarray}
S_2&=&\frac{\delta}{2^{1/2}3^{1/4}} { \mbox{ and  }}\nonumber  \\
u_1&=&\sqrt{3} v_1+3^{3/4}2^{-3/2}\delta-2^{3/2}3^{3/4}v_0^2.
\end{eqnarray}
Finally, to the next order $O\left( \varepsilon^3 \right) $, one gets
\begin{eqnarray}
\begin{bmatrix}
    0 & 0\\
    2/\sqrt{3}& 2
  \end{bmatrix}\left( \begin{array}{c}u_2\\
v_2
\end{array}\right)=
\begin{bmatrix}
    \beta{'}\frac{\partial ^{3}}{\partial \tau ^{3}} & \beta\frac{\partial ^{2}}{\partial \tau^{2}}-\frac{\partial^{4} }{\partial \tau ^{ 4}}\\
  \beta\frac{\partial ^{2} }{\partial \tau^{2}}-\frac{\partial^{4} }{\partial \tau ^{ 4}}   &  \beta{'}\frac{\partial ^{3}}{\partial \tau ^{3}} 
  \end{bmatrix}\left( \begin{array}{c}u_0\\
v_0
\end{array}\right) \nonumber \\ 
+
\left( \begin{array}{c}s_{3}\\
0
\end{array}\right)+
\left(\begin{array}{c} 4v_0^3\\
4\sqrt{3}\delta v_0-28\sqrt{3}v_0^3+8\sqrt{2}3^{1/4}v_0v_1
\end{array}\right). \nonumber 
\label{eq:fluc}
\end{eqnarray}
The solvability condition leads to 
\begin{equation}
\frac{\partial u_0}{\partial t}=S_3-4u_0^3+\delta u_0+\beta \frac{\partial^2 u_0}{\partial\tau^2}+\beta'\frac{\partial^3 u_0}{\partial\tau^3}-\frac{4}{3}\frac{\partial^4 v_0}{\partial\tau^4}.
\end{equation}
Introducing the following change of variables and parameters $\phi \equiv 2 u_0$,  $y\equiv 2 S_3$, 
and $c\equiv \delta$, we obtain the real generalized Swift-Hohenberg equation
\begin{equation}
\frac{\partial\phi}{\partial t}=y+c\phi-\phi^3+\beta\frac{\partial^2\phi}{\partial\tau^2}+\beta'\frac{\partial^3\phi}{\partial\tau^3}-\frac{4}{3}\frac{\partial^4\phi}{\partial\tau^4}.\label{eq:SH}
\end{equation}
Trivially, by normalizing the time $\tau$, and the disperson coefficients, we recover equation (\ref{Eq-1}). The Swift-Hohenberg equation is a well-known paradigm in the study of  pattern formation and the localized structures.  
Generically, it applies to systems that undergo  a symmetry breaking modulational  instability (often called Turing instability 
\cite{Lugiato1987}) close to the critical point associated with bistability (nascent optical bistability). It has been derived first under these conditions in hydrodynamics \cite{SH}, and later on in chemistry \cite{Hilali}, plant ecology \cite{Rene}, and nonlinear optics \cite{Mandel,Moloney}. Other real order parameter equations  in the form of nonvariational Swift-Hohenberg model have been also derived for spatially extended systems \cite{NonVariationalSG}. 

In the absence of the third order dispersion, the Swift-Hohenberg equation (\ref{eq:SH}) is variational, i.e., 
there exists a Lyapunov functional guaranteeing that evolution proceeds toward the state for which the 
functional has the smallest possible value which is compatible with the system boundary conditions. 
Without the third order dispersion, localized structures do not move. Indeed, these solutions are 
motionless and can become propagative if it is considered nonvariational terms 
in the dynamic \cite{AlejandroAlvarez2018}.
The conditions under which periodic patterns and localized structures appear are closely
related. Dynamically speaking, a sub-critical modulational instability
underlies the pinning phenomena responsible for the generation of temporal localized structures \cite{LS_SH}. 

The homogeneous steady states $\phi_s$ of Eq.~(\ref{eq:SH}) satisfies the cubic equation $y=\phi_s(\phi_s^2-c)$. The monostable regime will be for $c<0$, and the bistable regime will occur when $c>0$. The linear stability analysis with respect to finite frequency perturbation of the form $\phi=\phi_s+\delta \phi e^{\lambda t-i\omega\tau}$, with $\delta \phi\ll 1$,  yields the characteristic equation
\begin{equation}
\lambda=c-\beta\omega^2-i\beta'\omega^3-\frac{4}{3}\omega^4-3\phi_s^2
\label{eq:charSH}
\end{equation}
In the absence of the third order dispersion, i.e., $\beta'=0$, the homogeneous steady states undergo a modulational instability at
\begin{equation}
\phi_{M\pm}=\pm \sqrt{\frac{\beta^2}{16}+\frac{c}{3}} { \mbox{ and  }}  y_{M\pm}= \left(\frac{\beta^2}{16}-\frac{2c}{3}\right)\phi_{M\pm}
\label{thresholdSH}
\end{equation}
At both thresholds associated with MI, the critical frequency is $\omega_M^2=-3\beta/8$.
\section{Weakly nonlinear analysis}
\label{sec:weakana}
To calculate the nonlinear solutions bifurcating from the threshed associated with modulational instability, we use a weakly nonlinear analysis. 
To this end, we introduce an excess variable as $\phi \equiv \phi_s+\psi$.  We expand $\phi_s$, $\psi$ and $y$ in terms of a small parameter $\mu$ that measures the distance from the modulational instability threshold 
\begin{eqnarray}
\phi_s&=&\phi_{M\pm}+\mu \phi_{1}+\mu^{2} \phi_{2}+\mu^{3}  \phi_{3}+..., \nonumber \\
\psi&=&\mu \psi_{1}+\mu^2 \psi_{2}+\mu^3 \psi_{3}+..., \nonumber\\
y &=&y_{M\pm}+\mu y_{1}+\mu^2 y_{2}+\mu^3 y_{3}+....
\label{EXP}
\end{eqnarray}
The coordinates of the thresholds associated with the modulational instability $\phi_{M\pm}$ and $s_{M\pm}$ are explicitly given by Eq.~(\ref{thresholdSH}). We introduce a slow time
\begin{equation}
\frac{\partial }{\partial t}=\frac{\partial }{\partial t_1}+\mu^2\frac{\partial }{\partial t_2}.
\end{equation}
The solution to the homogeneous linear problem obtained at the leading order in $\mu$ is
\begin{equation}
\psi_{1}=W \exp{[i(\omega_M\tau+ \kappa t_1)]}+ cc,
\label{threshold}
\end{equation}
where $cc$ denotes the complex conjugate. The quantities $\psi_i$ and $\phi_{1}$ ($i = 1, 2$) can be calculated by inserting Eqs.~(\ref{EXP}) into the real Swift-Hohenberg Eq.~(\ref{eq:SH}) and equating terms with the same powers of $\mu$.

At third order in $\mu$, the solvability conditions yields the following amplitude equation
\begin{equation}
\partial A/\partial t =-6\alpha A+(f+ig)\left|A \right| ^{2}A,
\label{amp_eq}
\end{equation}
where $A=\mu W$, $\alpha=\phi_{s}-\phi_{M\pm}$ measures the distance from the second instability threshold and 
\begin{eqnarray}
f&=& \left[\frac{36\phi_{M\pm}^2(\frac{27}{8}\beta^2)}{\frac{729}{64}\beta^4+256 \kappa^2}+\frac{192}{\beta^2}\phi_{M\pm}^2 -3\right], \nonumber  \\
g&=&=\frac{576\phi_{M\pm}^2  \kappa}{\frac{729}{64}\beta^4+256 \kappa^2}.
\end{eqnarray}
The amplitude equation (\ref{amp_eq}) admits the following solution: $A = |A| \exp[i(\omega_0 t + \omega_M \tau)]$. The third order dispersion adds a new nonlinear phase $\omega_0$. By replacing this solution in Eq.~(\ref{amp_eq}), we obtain 
\begin{equation}
\left|A \right| ^{2}=6\frac{\phi_{s}-\phi_{M\pm}}{f},  { \mbox{ and  }} \omega_0=g\left|A \right| ^{2}.
\label{solu}
\end{equation}
\begin{figure}[h]
	\includegraphics[width=0.7\columnwidth]{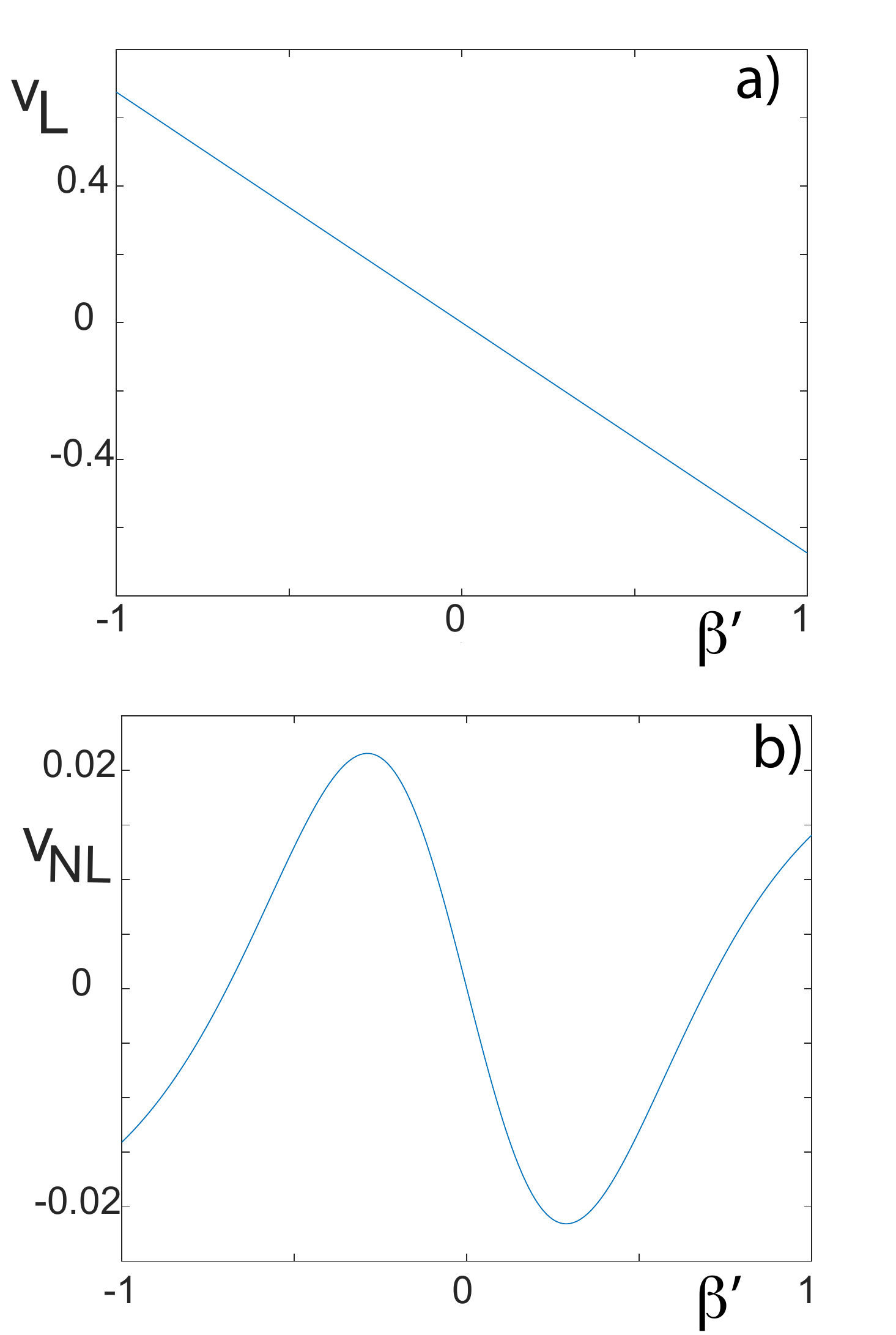}
	\caption{\label{fig:speed}Linear a) and nonlinear b) speed as a function of the parameter $\beta '$ computed from Eqs.~(\ref{eq:speeds}) for $\beta=-0.6$ and $c=-0.06$.
	}
\end{figure}
The nonlinear phase $\omega_0 $ is caused by the third order
dispersion. When taking into account the nonlinear
correction, the velocity takes the following form
\begin{equation}
v=v_{L}+v_{NL},
\end{equation}
where the linear and the nonlinear velocities are
 \begin{equation}
v_L=\frac{\partial Im(\lambda) }{\partial \omega_M}=-3\beta^\prime\omega^2  { \mbox{ and  }} v_{NL}=\frac{\partial \omega_0 }{\partial \omega_M}.
\label{eq:speeds}
\end{equation}
In the linear regime, the critical frequency, as well as the threshold associated with modulational instability, are not affected by the third order dispersion. The linear and nonlinear speeds as a function of the parameter $\beta '$ are shown in Fig.~(\ref{fig:speed}) for $\beta=-0.6$ and $c=-0.06$.

In the absence of the third order dispersion, the transition from super- to sub-critical modulational instability occurs when $c=c_{sub}=-87 \beta^2/38$ \cite{mous}. The modulational bifurcation is subcritical when $c>c_{sub}$. The sub-critical nature of the bifurcation can occur even in the monostable regime $c_{sub}<c<0$.

\section{Moving temporal localized structures}
\begin{figure}[h]
\includegraphics[width=0.95\columnwidth]{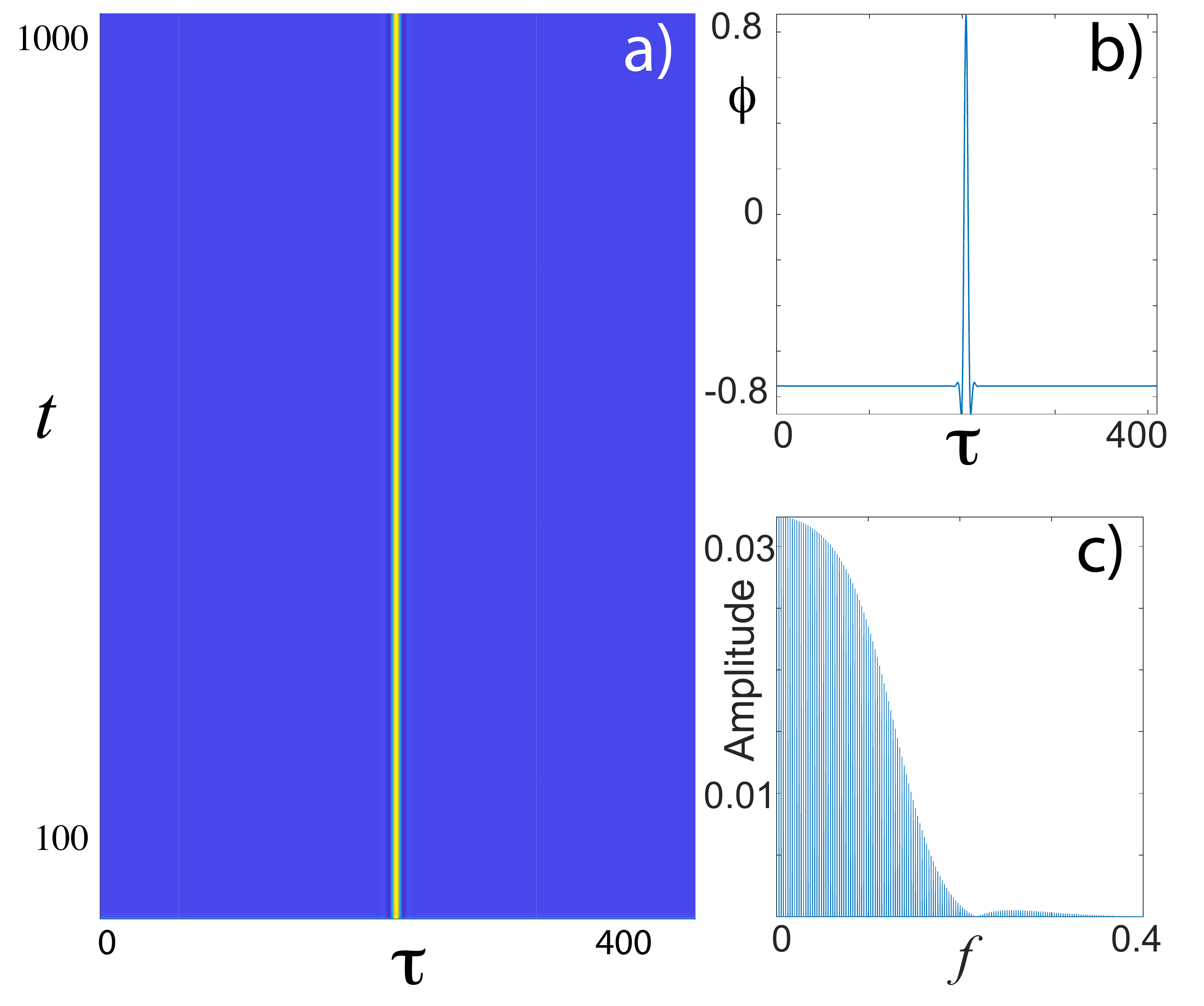}
\caption{\label{fig:profilesoli} Space time map a) and temporal profile b) of one localized structure, 
integrated for 1024 cells, $c=0.5$, $y=-0.05$, $\beta=-1$ and $\beta'=0$. c) shows the corresponding 
Fourier spectra, calculated using the integration of 500 roundtrips in the cavity.
}
\end{figure}

\begin{figure}[h]
	\includegraphics[width=0.9\columnwidth]{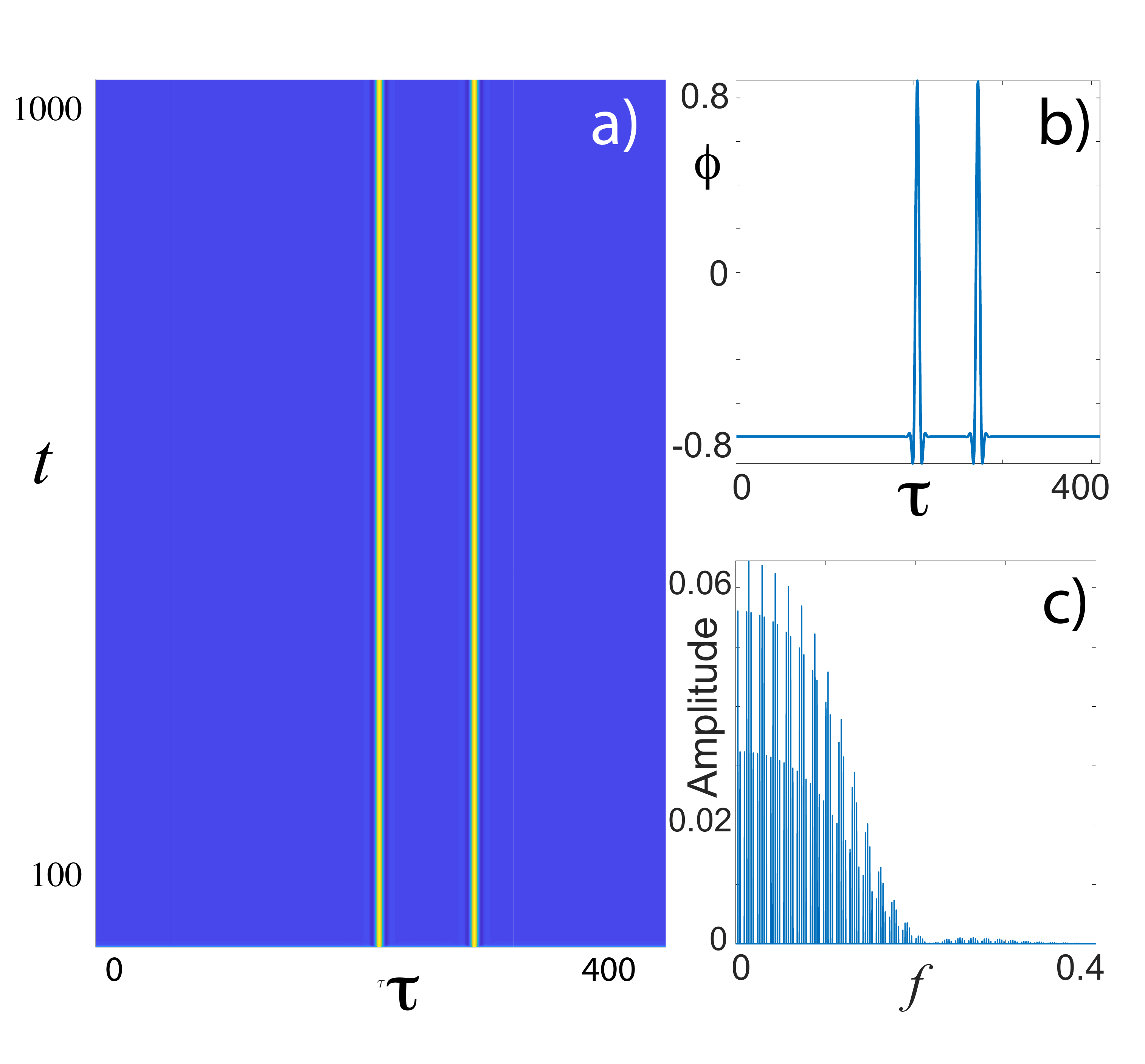}
	\caption{\label{fig:profilesoli2}Space time map a) and temporal profile b) of two localized structures, integrated for 1024 cells, $c=0.5$, $y=-0.05$, $\beta=-1$ and $\beta'=0$. c) shows the corresponding Fourier spectra, calculated using the integration of 500 roundtrips in the cavity.
	}
\end{figure}

\begin{figure}[h]
	\includegraphics[width=0.9\columnwidth]{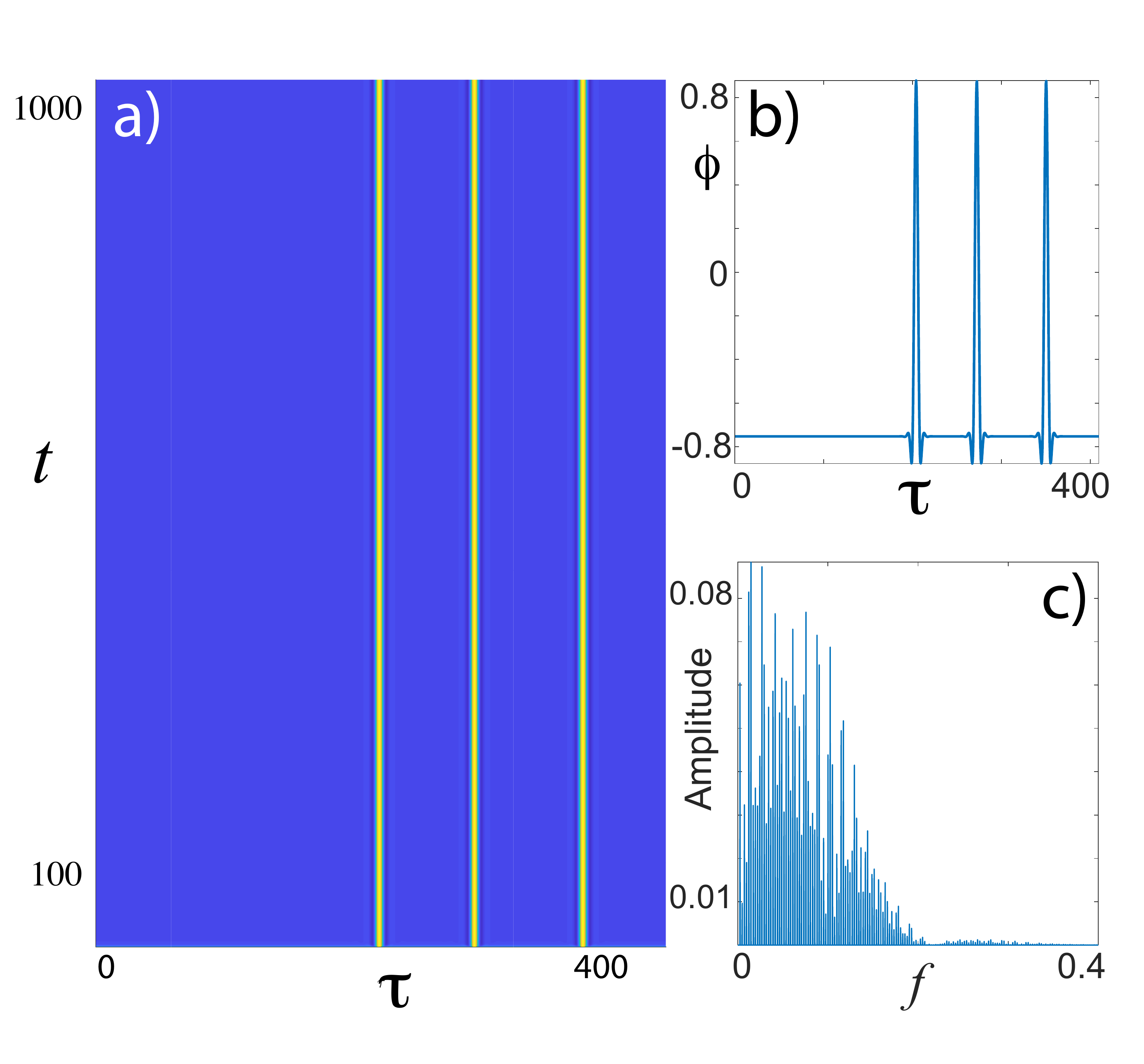}
	\caption{\label{fig:profilesoli3}Space time map a) and temporal profile 
	b) of three localized structures, integrated for 1024 cells, $c=0.5$, $y=-0.05$, $\beta=-1$ 
	and $\beta'=0$. c) shows the corresponding Fourier spectra, calculated using the integration of 500 roundtrips in the cavity.
	}
\end{figure}

\begin{figure}[h]
	\includegraphics[width=0.9\columnwidth]{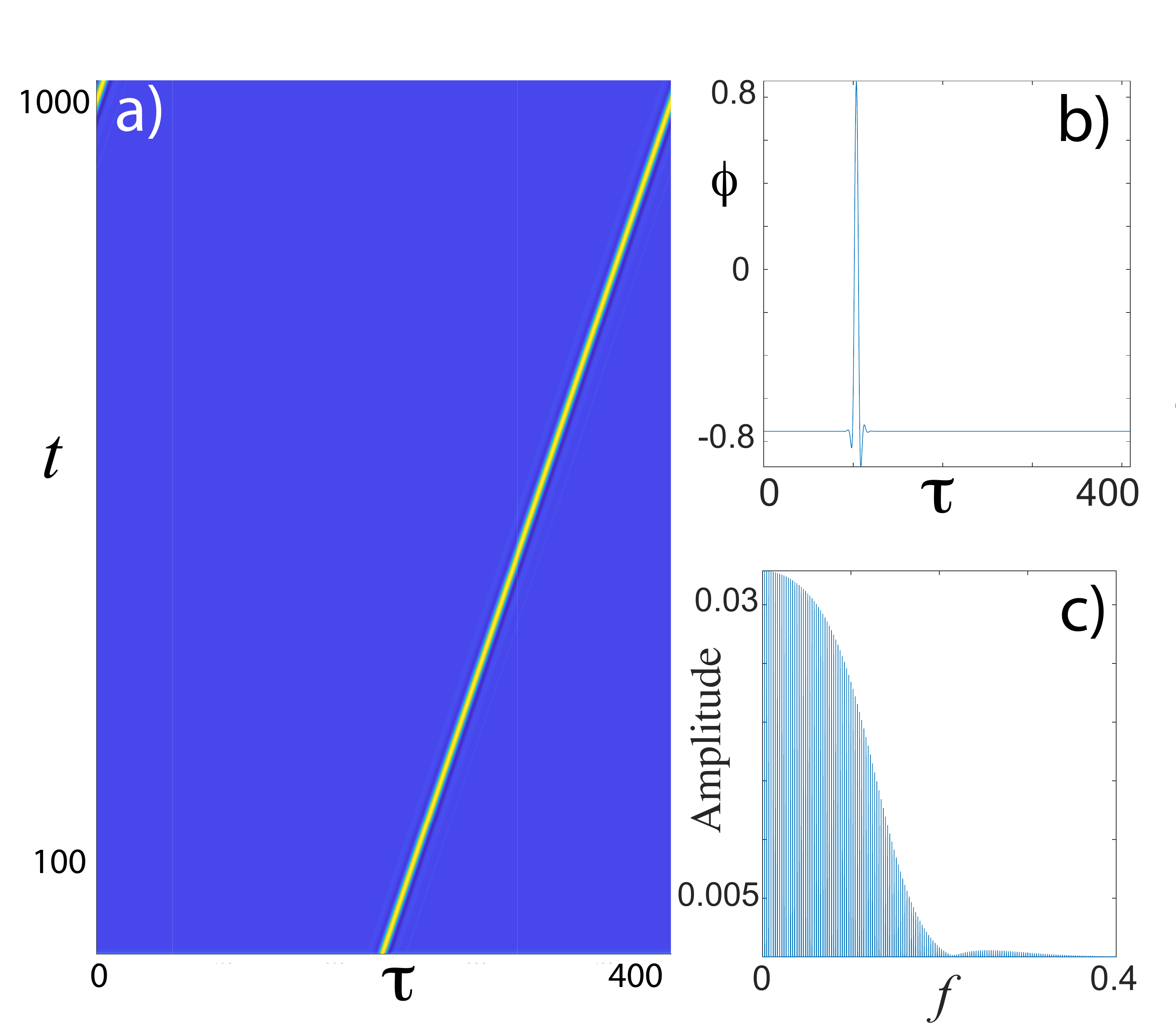}
	\caption{Space time map a) and temporal profile b) of one localized structure, integrated for 1024 cells, $c=0.5$, $y=-0.05$, $\beta=-1$ and $\beta'=0.5$. c) shows the corresponding Fourier spectra, calculated using the integration of 500 roundtrips in the cavity.
	}\label{fig:profilesolidyn}
\end{figure}

\begin{figure}[h]
	\includegraphics[width=0.9\columnwidth]{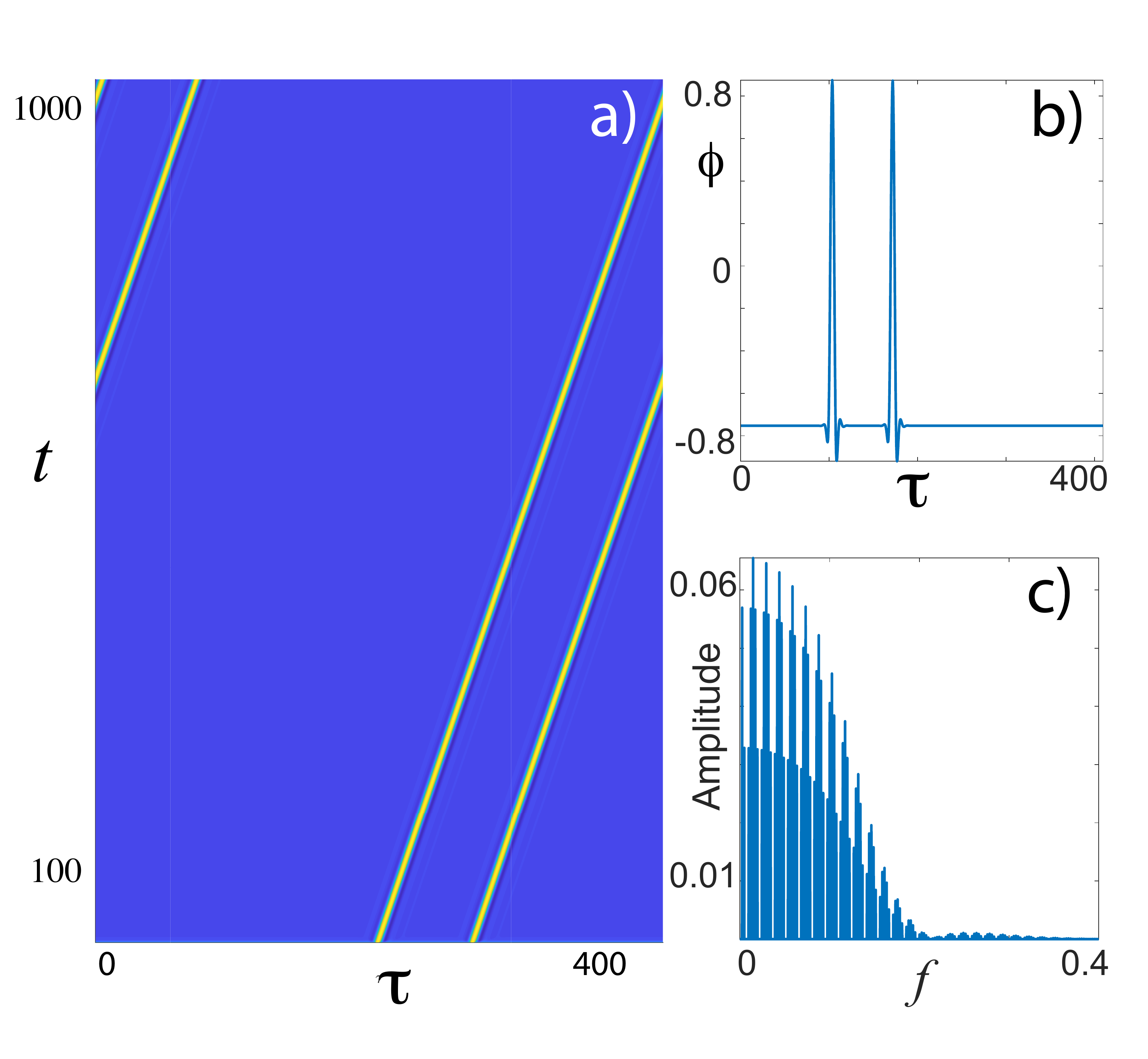}
	\caption{\label{fig:profilesolidyn2}Space time map a) and temporal profile b) of two localized structures, integrated for 1024 cells, $c=0.5$, $y=-0.05$, $\beta=-1$ and $\beta'=0.5$. c) shows the corresponding Fourier spectra, calculated using the integration of 500 roundtrips in the cavity.
	}
\end{figure}

\begin{figure}[h]
	\includegraphics[width=0.9\columnwidth]{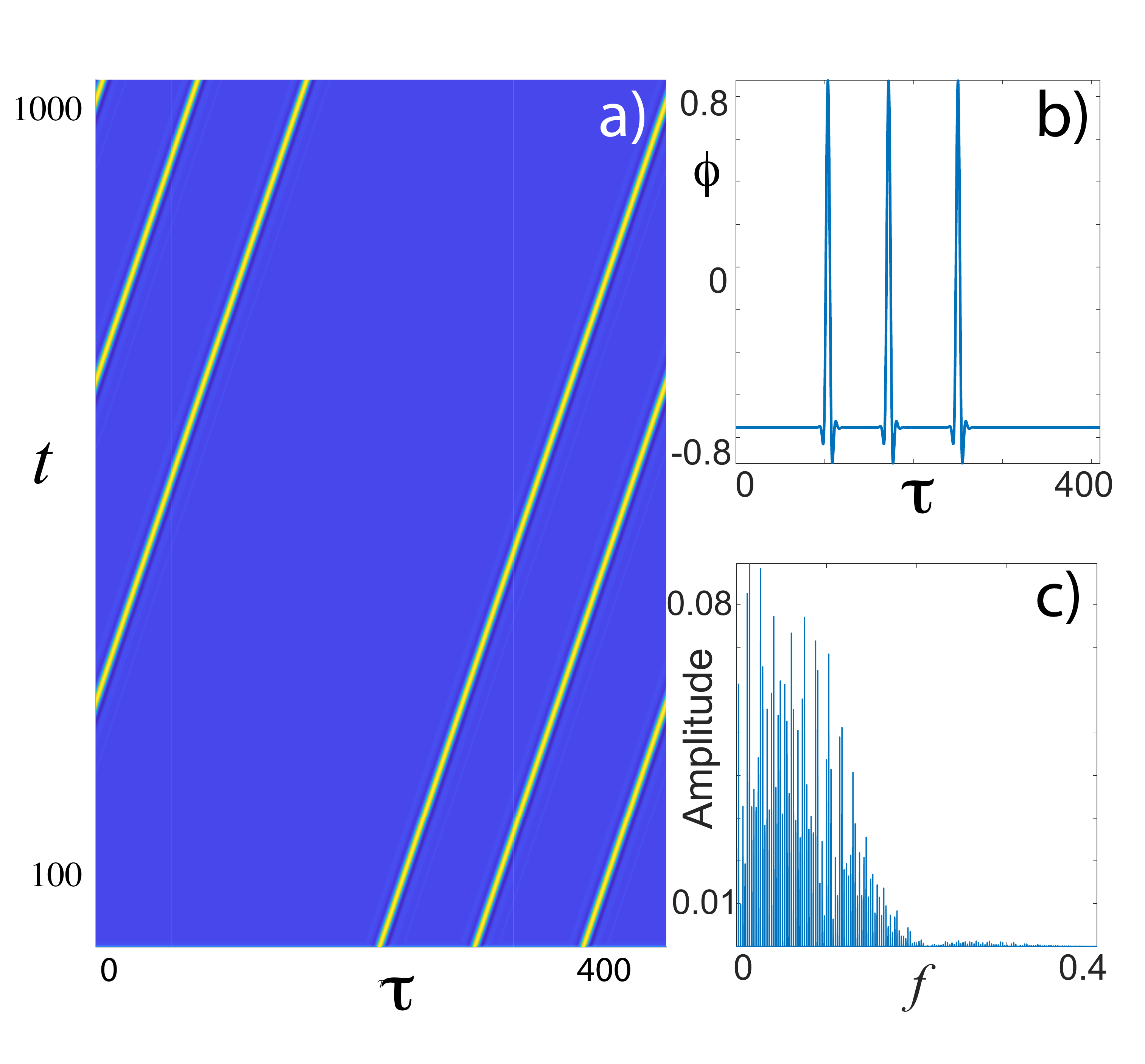}
	\caption{\label{fig:profilesolidyn3}Space time map a) and temporal profile b) of three localized structures, integrated for 1024 cells, $c=0.5$, $y=-0.05$, $\beta=-1$ and $\beta'=0.5$. c) shows the corresponding Fourier spectra, calculated using the integration of 500 roundtrips in the cavity.
	}
\end{figure}

Examples of a single, two or three peaks stationary symmetric temporal localized structures are shown in Figs.~(\ref{fig:profilesoli}, \ref{fig:profilesoli2}, and \ref{fig:profilesoli3}). They have been obtained numerically by using a periodic boundary condition compatible with the ring geometry of the optical resonator depicted in Fig.~(\ref{fig:LL}).  However, the third order dispersion breaks the reflection symmetry and allows for the motion of localized structures.  Examples of a single, two or three peaks moving temporal localized structures are shown in Figs.~(\ref{fig:profilesolidyn}, \ref{fig:profilesolidyn2}, and \ref{fig:profilesolidyn3}).    

Localized structures occur in the regime where the homogeneous steady state coexists with a spatially periodic structure. In addition, the system exhibits a high degree of multistability in a finite range of the control parameter values often called the pinning region \cite{Pomeau}. The number of LS and their temporal distribution along the longitudinal direction within the cavity is determined by the initial conditions. The interaction between the LS then leads to the formation of clusters or LS complexes. Dissipative structures have been observed in all areas of nonlinear science such as chemistry, biology, ecology, optics, and physics (see recent overview on this issue \cite{Reviews,Reviews2}.\\

\section{Pulses solutions for small third order dispersion}

\begin{figure}[b]
	\includegraphics[width=0.8\columnwidth]{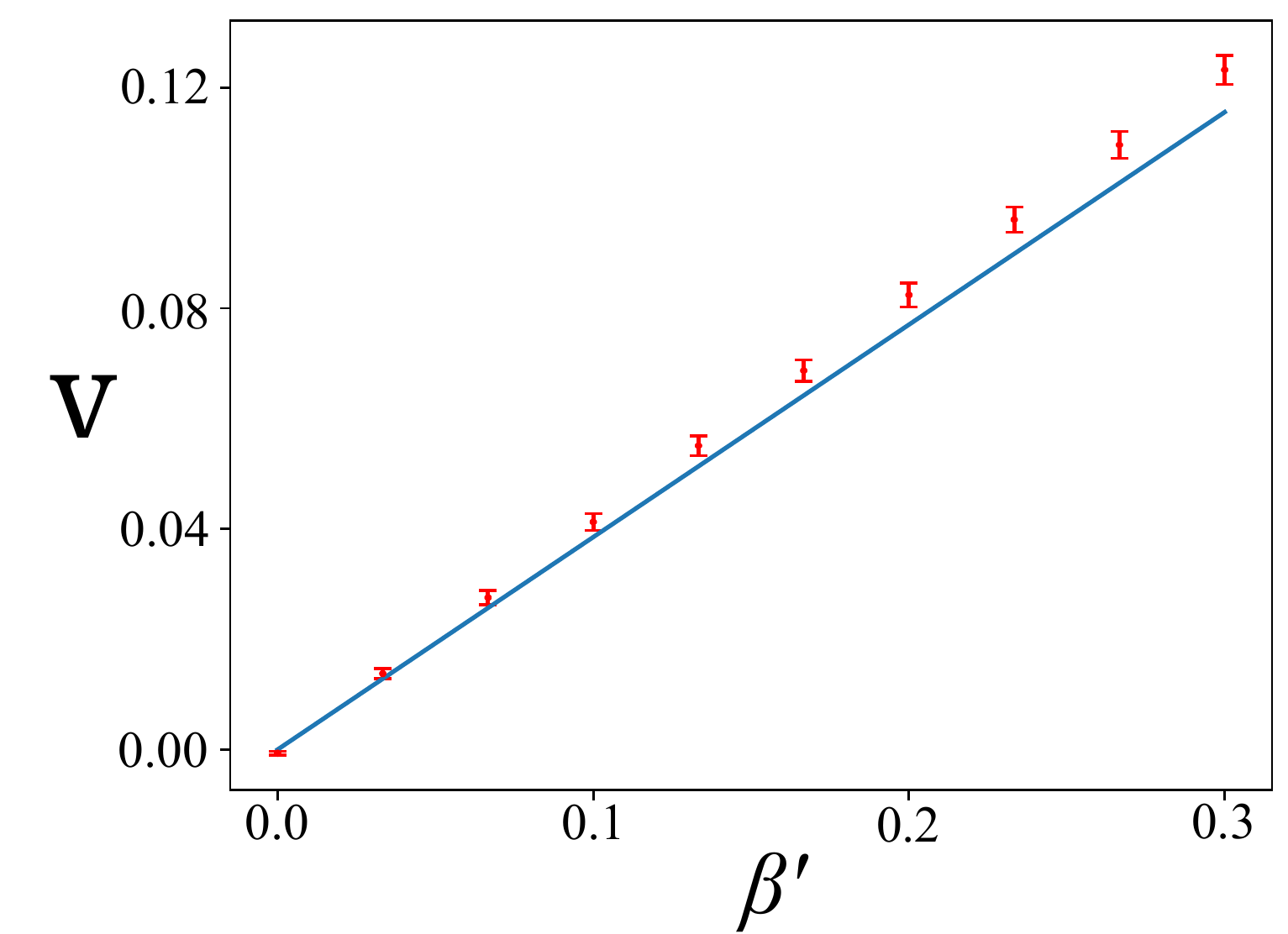}
	\caption{\label{fig:SpeedBeta} Localized structure speed as function of $\beta'$. Points are
	obtained from numerical simulations of Eq.~(\ref{eq:SH}). The continuous line corresponds to the plot of formula Eq.~(\ref{Eq-PulseSpeed}). 
	Parameters are $c=0.5$, $y=-0.05$, and $\beta=-1$.  
	}
\end{figure}

Another strategy to understand the effect of the third chromatic dispersion 
is to consider small $\beta'$.
Damped oscillations towards its flanks characterize the localized structure.
Analytical expressions of this motionless solution are not accessible as a consequence of the chaos theory \cite{DiegoContreras2015}.
Let us introduce $\phi_{0}(\tau-\tau_0)$ as the localized structure of model (\ref{eq:SH}) with $\beta' = 0$, where $\tau_0$
accounts for the temporal position of the global maximum of the localized structure.
Hence, the temporal variation of $\tau_0$ accounts for the speed of moving localized structure.
In order to calculate the speed LSs, we consider the following anstaz
\begin{equation}
\phi(t,\tau)=\phi_{0}(\tau-\tau_0(t))+w(t,\tau),
\label{Eq-AnsatzPulse}
\end{equation}
where $\tau_0$ is promoting to a temporal variable, which has variation 
of the order  $\beta'$ ($\dot{\tau}_0(t)\sim \beta'$) and $w(t,\tau)$ is a small correction function of the order of third chromatic dispersion.
Introducing the ansatz (\ref{Eq-AnsatzPulse}) in the generalized real Swift-Hohenberg equation (\ref{eq:SH}), linearizing in 
$w$, imposing the solvability conditions after straightforward calculations, we obtain
\begin{equation}
	\dot{\tau}_0=v\equiv \beta' \frac{\int (\frac{\partial^2 \phi_0}{\partial^2 \tau})^2   d\tau}{\int \left(\frac{\partial \phi_0}{\partial \tau}\right)^2   d\tau}.
\label{Eq-PulseSpeed}
\end{equation}
Note that the speed of propagation of the pulse is proportional to the third chromatic dispersion.
From Eq.~(\ref{Eq-PulseSpeed}) we see that, if $\beta'$ is positive (negative), the localized structure 
propagates towards the positive (negative) flank.
Figure~(\ref{fig:SpeedBeta}) shows a comparison between analytical expression of the speed of single peak LS, Eq.~ (\ref{Eq-PulseSpeed}), 
and numerical simulations of the governing equation (\ref{eq:SH}). From this figure, we can infer a quite good agreement when $\beta'$ is small.

\section{Conclusions}
Employing a multiple-scale reduction near the instability threshold associated with the modulational instability 
and close to a second-order critical point marking the onset of a hysteresis loop (nascent bistability), 
we have derived a real Swift-Hohenberg equation describing the evolution of the envelope of the electric 
field circulating inside an optical Kerr resonator. The presence of third-order dispersion renders 
the obtained Swift-Hohenberg equation nonvariational meaning there is no free energy or 
Lyapunov functional to minimize.
 A linear and a weakly nonlinear analysis have been performed 
to identify the conditions under which a transition from super- to sub-critical modulational instability takes place.  
More importantly, we have shown that the third order dispersion allows for temporal localized structures 
to move with a constant speed as a result of the broken reflexion symmetry.  
We have characterized this motion by estimating the speed associated with it.

\section*{Acknowledgments}
K.P. acknowledges the Methusalem foundation for financial support. 
M.T. is a Research Director with the Fonds de la Recherche Scientifique F.R.S.-FNRS, Belgium. 
MGC and MF thank financial support of  the Millennium Institute for Research in Optics (Miro).

\end{document}